\documentclass[prl,twocolumn]{revtex4}
\usepackage{amsmath}
\usepackage{graphics}

\newcommand \be{\begin{equation}}
\newcommand \ee{\end{equation}}
\begin{document}

\title{\textbf{The electroweak theory with \textit{a priori} superluminal
neutrinos and its physical consequences}}
\author{C.A. Dartora$^{1}$\thanks{%
cadartora@eletrica.ufpr.br} \ and \ G.G. Cabrera$^{2}$\thanks{%
cabrera@ifi.unicamp.br}}
\affiliation{$^{1}$Electrical Engineering Department, Federal University of Parana
(UFPR), Brazil\\
$^{2}$ Instituto de F\'{\i}sica \ \ `Gleb Wataghin', Universidade Estadual
de Campinas (UNICAMP), C.P. 6165, Campinas 13.083-970 SP, Brazil\\
}

\begin{abstract}
{In recent experiments conducted by the OPERA collaboration, researchers
claimed the observation of neutrinos propagating faster than the light speed
in vacuum. If correct, their results raise several issues concerning the
special theory of relativity and the standard model of fundamental
particles. Here, the physical consequences of superluminal neutrinos
described by a tachyonic Dirac lagrangian, are explored within the standard
model of electroweak interactions. If neutrino tachyonic behavior is
allowed, it could provide a simple explanation for the parity violation in
weak interactions and why electroweak theory has a chiral aspect, leading to
invariance under a $SU_{L}(2)\times U_{Y}(1)$ gauge group. Right-handed
neutrino becomes sterile and decoupled from the other particles quite
naturally.}

\end{abstract}

\maketitle

\section{Introduction}

The OPERA collaboration \cite{[OPERA]} recently announced the astonishing
results obtained in a long baseline neutrino experiment designed to measure,
among other things, the neutrino time-of-flight with the best currently
known estimation of time and distance using atomic clocks and GPS system.
They claim that the neutrinos with energies in the range $10-60$GeV traveled
the approximate $730$km from the CERN facilities to the Gran Sasso detector
faster than light would do, with $(v-c)/c=2.5\times 10^{-5}$ within a
six-sigma level of confidence. The result more or less corroborate an
earlier experimental finding conducted by the MINOS collaboration \cite%
{[MINOS]}, but clearly contradicts astronomical available data. Measurement
of neutrino with energies in the range of a few MeV coming from supernova
SN1987A put the stringent limit of $(v-c)/c<10^{-9}$ \cite%
{[SN1987A],[Fagion]}. As a matter of fact, the sign of the muon-neutrino
squared mass $m_{\nu }^{2}$ is still open to debate \cite%
{[caban],[eidel],[assama]}, and its implications were discussed in Ref. \cite%
{[erlich]}. In favor of the neutrino superluminal behavior, in Ref. \cite%
{[JJ]} J. Ciborowski and J. Rembielinski proposed that tachyonic neutrinos
could explain certain anomalies in the decay of tritium \cite{[tritium]}. On
the other hand, serious constraints on the existence of superluminal
neutrinos were established by Cohen and Glashow \cite{[CG]}, arguing that if
the standard model as we know still applies, such neutrinos should lose
energy by producing photons and $e^{+}e^{-}$ pairs through $Z_{0}$ mediated
processes, analogous to a bremsstrahlung radiation effect, which would alter
the energy spectrum of detected neutrinos. The search of pair production by
the ICARUS Collaboration found no evidence of anomalous $e^{+}e^{-}$ pair
production\cite{[ICARUS]}.

Currently, the OPERA result still waits to be confirmed or refuted by
independent experiments, but if correct we are facing a ground-breaking
moment in Physics comparable to that of the Michelson-Morley experiment. At
this point we can only speculate on a few number of hypotheses to explain
the observed data: i) The OPERA experimental results are not correct due to
the presence of a subtle systematic error which is not being taken into
account yet. Attempts to detect possible sources of such errors are being
done, such as in Ref. \cite{[cern1]}, which posed the question on the
influence of data filtering, but the OPERA team already refuted those
objections conducting a series of improved experiments; ii) The neutrinos
travel faster than light but are not tachyonic in character, which would
mean that the true value of $c$ appearing in the special theory of
relativity is the neutrino speed. As a matter of fact, one can speculate
that since the photon is far more interacting than the neutrino the light
speed in a quantum-mechanical vacuum is a renormalized value; iii) The
neutrinos are tachyons and travel faster than light. The idea of tachyonic
neutrinos are not new\cite{[JJ],[TN]}, but there are a number of arguments
against it \cite{[ATN]}. In spite of the above, if we accept that the
special theory of relativity remains valid including anti-orthochronous
Lorentz transformations describing particles with $v>c$, this fact has to to
be taken into account in the standard model of particle physics; iv) The
special theory relativity in four-dimensional spacetime reached its limit of
validity and a profound reformulation of the physical theories are in order.
For instance we may cite the possibility of theories having extra spacetime
dimensions \cite{[dim]}.

The purpose of the present paper is to explore the physical consequences of
having \textit{a priori} superluminal neutrinos in the standard model of
elementary particles. We assume that the special theory of relativity is
rigorously valid and consider that a complete physical theory must include
the superluminal sector of the full Lorentz-Poincar\'{e} group, the
so-called anti-orthochronous Lorentz transformations, in the form of a
tachyonic Dirac equation. A particular interpretation of the special
relativity says that massive particles may exist in states with $c>v$
(subluminal or bradyonic particles) or $v>c$ (superluminal or tachyonic
particles) \cite{[recami]} but the frontier dividing subluminal and
superluminal motion cannot be crossed, i.e., a bradyon cannot be converted
classically into a tachyon and vice-versa. The standard model extended to
include the tachyonic sector starts from a lagrangian density describing two
Dirac fields, but one of them, identified with the neutrinos, is tachyonic.
As we will show in the following Section, the Dirac equation describing
superluminal fermions slightly differs from the subluminal Dirac equation
but leads to important physical consequences.


\section{The Dirac equation for bradyons and tachyons}

In the present Section we briefly review the fundamental aspects of the
Dirac equation in the bradyonic and tachyonic forms. The special theory of
relativity requires the invariance of the quantity $p_\mu p^\mu = m^2$,
which is a Casimir invariant of the Lorentz-Poincar\'e group, where $%
\mu=0,1,2,3$ are the spacetime indices, $p^\mu = (E,\mathbf{p})$ is the
four-momentum vector, $E$ is the energy and $\mathbf{p}$ is the linear
momentum and $m$ is the rest mass of the particle. The Einstein convention
of summing up the repeated indices is implied throughout this paper.

For a particle with rest mass $m^{2}>0$ the dispersion relation can be
written as $E^{2}=\mathbf{p}^{2}+m^{2}$, and the motion is subluminal ($v<c$%
). The Dirac equation can be directly obtained from $E=\sqrt{\mathbf{p}%
^{2}+m^{2}}$. This procedure is now standard and found in textbooks \cite%
{[sakurai],[greiner],[weinberg],[ryder]}. For spin $1/2$ fermions we get: 
\begin{equation}
(i\gamma ^{\mu }\partial _{\mu }-m)\psi =0~,  \label{eqDirac1}
\end{equation}%
being $\gamma ^{\mu }$ the Dirac gamma matrices and $\psi $ a four component
Dirac spinor. The above equation is invariant under the symmetries of charge
conjugation $C$, parity $P$ (corresponds change the sign of the spatial
coordinates, i.e., $\mathbf{x}\rightarrow -\mathbf{x}$) and time reversal $T$
(corresponds to reverse the time flow, $t\rightarrow -t$), separately and
also the product $CPT$\cite{[sakurai],[greiner],[weinberg],[ryder]}. The
Dirac equation can be obtained via Euler-Lagrange equations from the
following lagrangian density for a massive Dirac field: 
\begin{equation}
\mathcal{L}=i\bar{\psi}\gamma ^{\mu }\partial _{\mu }\psi -m\bar{\psi}\psi ~.
\label{lagfreeSub}
\end{equation}

The tachyonic dispersion relation, for which $v>c$, is obtained by making
the replacement $m\rightarrow \pm im$, meaning that a superluminal particle
has imaginary mass and $m^{2}<0$. This way the relation $E^{2}=p^{2}-m^{2}$
leads to the corresponding tachyonic Dirac equation, which was obtained in
Ref. \cite{[TacDirac1]}. The result is: 
\begin{equation}
(i\gamma ^{5}\gamma ^{\mu }\partial _{\mu }-m)\psi =0~,  \label{eqDiracty}
\end{equation}%
where $\gamma ^{5}=i\gamma ^{0}\gamma ^{1}\gamma ^{2}\gamma ^{3}$ is the
well known chirality operator. 
The detailed analysis of $CPT$ invariance was performed in Ref. \cite%
{[TacDirac2]}, showing that the tachyonic Dirac equation is invariant under
the product $CP$ and $T$ separately, but not under $C$ or $P$ alone. Such a
property of tachyonic Dirac equation could explain why the electroweak
interactions violate parity symmetry $P$, if neutrinos are tachyonic
particles. The lagrangian density for a tachyonic Dirac field is written
below for completeness: 
\begin{equation}
\mathcal{L}=i\bar{\psi}\gamma ^{5}\gamma ^{\mu }\partial _{\mu }\psi -m\bar{%
\psi}\psi ~,  \label{lagfreeSuper}
\end{equation}%
where $m$ is now a real parameter corresponding to the particle's rest mass.

Notice that both the subluminal and tachyonic Dirac fields describe a
luminal particle with $v=c$ in the limit $m\rightarrow 0$, but the
convergence to the speed limit $v=c$ is obtained from the left for the case
of massive field with $v<c$, and from the right for the tachyonic massive
field with $v>c$.

\section{The Electroweak Theory with A Priori Superluminal Neutrino}

Our starting point is a lagrangian density describing two massless Dirac
fields written below: 
\begin{equation}
\mathcal{L}=i\bar{\psi}_{e}\gamma ^{\mu }\partial _{\mu }\psi _{e}+i\bar{\psi%
}_{\nu }\gamma ^{5}\gamma ^{\mu }\partial _{\mu }\psi _{\nu }~,
\label{lagfree1}
\end{equation}%
where $\psi _{e}$ and $\psi _{\nu }$ describe the electron and the
neutrino-electron field respectively. Notice that the electron becomes
subluminal if a mass $m>0$ is attributed to the field $\psi _{e}$. By
contrast, the neutrino field becomes superluminal in the case $m>0$. For the
sake of convenience, the fields can be split up into chirality eigenstates,
using the projection operators $\Pi _{\pm }=(1\pm \gamma ^{5})/2$, yielding: 
\begin{equation*}
\psi _{eL}=\frac{1-\gamma ^{5}}{2}\psi _{e}~,~\psi _{\nu L}=\frac{1-\gamma
^{5}}{2}\psi _{\nu }~~,~~
\end{equation*}%
\begin{equation*}
\psi _{eR}=\frac{1+\gamma ^{5}}{2}\psi _{e}~,~\psi _{\nu R}=\frac{1+\gamma
^{5}}{2}\psi _{\nu }~~,~~
\end{equation*}%
where the subscript $L(R)$ are customary known as left(right)-handed
particles, but in fact they correspond to negative(positive) chirality
eigenvalues. Notice that for idempotent operators $\Pi _{\pm }^{2}=\Pi _{\pm
}$ and for any Dirac spinor $\psi $ we have $\psi =(\Pi _{+}+\Pi _{-})\psi $%
, allowing to straightforwardly rewrite expression (\ref{lagfree1}) as
follows: 
\begin{eqnarray}
\mathcal{L} &=&i\bar{\psi}_{eL}\gamma ^{\mu }\partial _{\mu }\psi _{eL}+i%
\bar{\psi}_{\nu L}\gamma ^{\mu }\partial _{\mu }\psi _{\nu L}+  \notag \\
&&i\bar{\psi}_{eR}\gamma ^{\mu }\partial _{\mu }\psi _{eR}-i\bar{\psi}_{\nu
R}\gamma ^{\mu }\partial _{\mu }\psi _{\nu R}~.  \label{lagfree2}
\end{eqnarray}%
The above equation shows that the superluminal neutrino field ends up with a
sign difference between the left and right chirality components, which is
not the case of electrons. In order to appreciate the significance of such a
fact, we merge the electron and neutrino fields into left-handed and
right-handed isodoublets: 
\begin{eqnarray}
\Psi _{L} &=&\left( 
\begin{array}{c}
\psi _{eL} \\ 
\psi _{\nu L}%
\end{array}%
\right) ~~,  \label{LMerged} \\
\Psi _{R} &=&\left( 
\begin{array}{c}
\psi _{eR} \\ 
\psi _{\nu R}%
\end{array}%
\right) ~~,  \label{RMerged}
\end{eqnarray}%
enabling us to recast equation (\ref{lagfree2}) into the following form: 
\begin{equation}
\mathcal{L}=i\bar{\Psi}_{L}\gamma ^{\mu }\partial _{\mu }\Psi _{L}+i\bar{\Psi%
}_{R}\gamma ^{\mu }\partial _{\mu }\tau _{z}\Psi _{R}~,  \label{lagfreeLR}
\end{equation}%
where $\tau _{z}$ is the Pauli matrix 
\begin{equation*}
\tau _{z}=\left( 
\begin{array}{cc}
1 & 0 \\ 
0 & -1%
\end{array}%
\right) ~,
\end{equation*}%
which operates in the so-called weak isospin space, yielding 
\begin{equation*}
\tau _{z}\Psi _{R}=\left( 
\begin{array}{c}
\psi _{eR} \\ 
-\psi _{\nu R}%
\end{array}%
\right) ~.
\end{equation*}

Now, we can straightforwardly construct a gauge theory of electroweak
interactions, requiring gauge invariance of the above lagrangian density\cite%
{[weinberg],[ryder]}. For the lagrangian density describing the left-handed
isodoublet, $\mathcal{L}_{L}=i\bar{\Psi}_{L}\gamma ^{\mu }\partial _{\mu
}\Psi _{L}$, we require invariance under a general transformation of the
gauge group $SU_{L}(2)\times U_{Y}(1)$: 
\begin{equation*}
\Psi _{L}^{\prime }=\exp \left[ i\Lambda _{0}+i\frac{\boldsymbol{\tau }\cdot 
\boldsymbol{\Lambda }}{2}\right] \Psi _{L}~,
\end{equation*}%
where $\Lambda _{0}$ and $\boldsymbol{\Lambda }$ are well-behaved spacetime
functions and $\boldsymbol{\tau }$ are the Pauli isospin matrices. This is
achieved by replacing ordinary derivatives $\partial _{\mu }$ by covariant
ones: 
\begin{equation}
D_{\mu L}=\partial _{\mu }+ig^{\prime }X_{\mu }-ig\boldsymbol{\tau }\cdot 
\mathbf{W}_{\mu }~,  \label{covleft}
\end{equation}%
where $X_{\mu }$ and $\mathbf{W}_{\mu }$ are the $U_{Y}(1)$ and $SU_{L}(2)$
gauge fields, respectively, and $g^{\prime }$ and $g$ are the corresponding
couplings. The $SU_{L}(2)$ gauge group could be interpreted as a \ local 
\emph{`rotation'}  mixing of the subluminal and superluminal spinor fields, $%
i.e.$, connecting the orthochronous to the anti-orthochronous Lorentz groups
in the limit $v\rightarrow c$. By contrast, the lagrangian density
representing the right-handed isodoublet, $\mathcal{L}_{R}=i\bar{\Psi}%
_{R}\gamma ^{\mu }\partial _{\mu }\tau _{z}\Psi _{R}$, is not invariant
under general $SU_{R}(2)$ \ `rotations' due to the presence of the $\tau _{z}
$ matrix. Only rotations around the $z$-axis of the weak isospin would
preserve the gauge invariance of the right-handed sector. This way, we can
break the right-handed isodoublet into two isosinglets $\psi _{eR}$ and $%
\psi _{\nu R}$ by postulating a special case of $U_{Y}(1)$ gauge
transformation connected to the weak hypercharge $Y$ as follows: 
\begin{equation*}
\Psi _{R}^{\prime }=\exp \left[ i(1+\tau _{z})\Lambda _{0}\right] \Psi _{R}~,
\end{equation*}%
where $\Lambda _{0}$ is a gauge change related to the $U_{Y}(1)$ gauge
group. The covariant derivative of the right-handed sector is written as: 
\begin{equation}
D_{\mu R}=\partial _{\mu }+ig^{\prime }(1+\tau _{z})X_{\mu },
\label{covright}
\end{equation}%
allowing us to obtain the usual Weinberg-Glashow-Salam electroweak theory,
described in more details elsewhere\cite{[weinberg],[ryder]}. The Higgs
sector of the theory, which provides mass to the physical fields via
symmetry breaking, can be introduced in the next stage, making the neutrinos
superluminal, $m^{2}<0$. Notice that the coupling constant of the
right-handed sector of the theory to the $SU_{R}(2)$ gauge field $\mathbf{W}%
_{\mu }$ vanishes in order to preserve the gauge invariance of the theory, $%
i.e.$, the right-handed particles do not transform as an isodoublet. The
right-handed neutrino is also decoupled from the gauge field $X_{\mu }$,
thus becoming \ `sterile'.

As a final remark, we mention that if the fields acquire mass the chirality
eigenstates get mixed up \cite{[ryder]} ($-m\bar{\psi}_{\nu }\psi _{\nu }=-m[%
\bar{\psi}_{\nu L}\psi _{\nu R}+\bar{\psi}_{\nu R}\psi _{\nu L}]$), allowing
the conversion of a left-handed neutrino into a right-handed one, despite
the fact that only the left-handed eigenstate is physically observable
through the electroweak interactions. This is a new kind of neutrino
oscillation which mixes left and right-handed neutrinos of the same flavor
(the other kind of neutrino oscillation being the well-known neutrino flavor
oscillation \cite{[nosc]}). The mean velocity of the neutrino field is given
by: 
\begin{equation}
\left\langle \frac{d\mathbf{x}}{dt}\right\rangle =\int d^{3}\mathbf{x\ }\bar{%
\psi}_{\nu }\gamma ^{5}\mathbf{\gamma }\psi _{\nu }~,  \label{meanv}
\end{equation}%
where $\mathbf{\gamma }=(\gamma ^{1},\gamma ^{2},\gamma ^{3})$ is the vector
containing the spatial components of the Dirac gamma matrices. Writing the
above expression in terms of chirality eigenstates, we obtain: 
\begin{equation}
\left\langle \frac{d\mathbf{x}}{dt}\right\rangle =\int d^{3}\mathbf{x}[\bar{%
\psi}_{\nu L}\mathbf{\gamma }\psi _{\nu L}-\bar{\psi}_{\nu R}\mathbf{\gamma }%
\psi _{\nu R}]~.  \label{meanv1}
\end{equation}%
Looking at the above expression, the mean velocity of the neutrino field
depends on both left and right-handed components. Even if the physically
observable left-handed neutrino field $\psi _{\nu L}$ becomes subluminal
when it acquires mass, the quantum interference effect due to the mixing of
the left-handed and the sterile right-handed neutrinos provided by the mass
term, would allow for a superluminal velocity, which could depend on the
propagation distance of the neutrino beam.

\section{Conclusion}

In summary, we discussed the possibility of existence and its physical
consequences of an \textit{a priori} tachyonic neutrino in the standard
model of electroweak interactions. Despite the conceptual difficulties
introduced by the presence of superluminal particles in physical theories,
as well as the stringent limits imposed on such behavior by experimental
observations, the existence of a tachyonic neutrino, if allowed and not
ruled out by future experiments, would explain in a \ `natural' way the
parity violation, the existence of chiral gauge transformations and the
decoupling of the right-handed neutrino, which becomes sterile, from the
electroweak field.

\end{document}